\documentstyle[preprint,aps]{revtex}
\begin{document}
\draft

\title{Localization of a Polymer with Internal Constraints}
\author{J. D. Bryngelson}
\address{
Division~of~Computer~Research~and~Technology, \\
National Institutes of Health, Bethesda, MD 20892}
\author{D. Thirumalai}
\address{
Institute~for~Physical~Science~and~Technology \\
University~of~Maryland, College~Park, MD 20742}
\date{               }
\maketitle
\begin{abstract}
We argue that Kantor and Kardar's assertion that their simulation
results contradict our criterion for the localization of a softly constrained
ideal polymer is incorrect.  Our criterion is inapplicable to the model
used in these simulations.  
We argue that our criterion for localization of as polymer with internal
constraints strongly suggests that the models with soft and hard constraints
belong to different universality classes.
The size measure used by Kantor and Kardar 
is also not adequate for studying polymer localization.  We also show
that, in spite of these contrasts, combining the simulation results with our
criterion for ideal chains seems to uncover some intriguing physics.
\end{abstract}
\pacs{PACS numbers: 82.35.+t, 87.15.By, 61.41.+t }

In the accompanying comment, Kantor and Kardar (KK) ~\cite{KK}
discuss simulations 
of an ideal polymer chain of $N$ monomers 
with $M$ randomly chosen distance constraints.
These simulations use an analogy between ideal polymers and resistor 
networks ~\cite{resistor} to 
calculate the mean-squared end-to-end distance
$(r^2)$ of the polymer.  
The simulations together with a lower bound for $r^2$ are used to suggest that
the polymer remains extended unless the number of constraints exceeds
some $M_c$ which is of order $N$.
They concluded that this result is at variance with our recent
theory in which we argue that $M_c \sim N/\log N$
for an {\em entirely different} model (BT) ~\cite{us}.
Here we show that
this assertion is too strong, and that their results for 
ideal polymers may shed
an interesting light on ours.

The most obvious contrast between our study and that of KK is that the
two studies investigate different models.  
In the random resistor analogy a short
corresponds to a Dirac delta function (a hard constraint).  In our Letter we
{\em explicitly excluded} such delta function constraints; see footnote 9 
of BT.  Although a polymer with soft constraints will be more expanded
than one with hard constraints, 
soft constraints add a new length scale (inducing longer range interactions)
to the system
which is present in our localization condition
(see equations 1 and 10 of BT).
The size of a polymer with soft constraints depends on
{\em both} the number of constraints and the size of the constraints.
Thus, the two models may be in different universality classes, making
the objections of KK circumspect.  
The above notwithstanding, 
expanding our localization condition (eq. 10 of BT)
for large $\phi \equiv (b/c)^2$ 
and small $\delta \equiv 1-\mu$ gives the localization condition
\begin{equation}
\delta < \frac{\log {\cal L} - \log 2}{\log \phi + \log {\cal L} -1}
       + {\rm O}(\frac{1}{\phi}) + {\rm O}(\delta^2).
\end{equation}
The hard constraint case corresponds to the limit $\phi \rightarrow \infty $
with $\phi \gg {\cal L} \gg 1$, yielding a maximum value of 
$\delta \rightarrow 0$, 
self-consistent with our expansion and consistent (!) with the arguments of KK
that the critical value of $M \sim N$ in this limit.

The purpose of our Letter was to determine whether a polymer with soft internal
constraints is better approximated by an expanded chain or by a collapsed chain.
The nature of the transition between these two regimes could not be 
determined by our methods .
A smooth, continuous transition, as found in the simulations of KK is 
not inconsistent with our results.  The question then becomes, what is
the size of the polymer at this transition?  
If we {\em assume} that 
$R_G^2 \propto N/M$ holds for the radius of gyration ($R_G$) of our model
~\cite{BP,SV},
then at the transition between the two regimes, $R_G^2 \sim \log N$.
This scaling is a reasonable dividing
line between the fully expanded ($R_G^2 \sim N$) and fully collapsed
($R_G^2 \sim O(1)$) regimes.  Furthermore, this dividing line is {\em not}
arbitrary.  It is consistent with a completely different criterion for
polymer localization, as we now show.  
For a general system of $N$ (ideal) monomers connected by harmonic 
bonds, 
\begin{equation}
R_G^2 \sim \frac{1}{N} \sum_{i=1}^{N-1}\frac{1}{\omega_i}
      \sim \int_{\omega_{min}}^{\omega_{max}} \frac{n(\omega)}{\omega} d\omega,
\end{equation}
where the $\omega_i$ are the non-zero eigenvalues of the connectivity matrix,
$\omega_{min}$ is the lowest non-zero eigenvalue, $\omega_{max}$ is the 
largest eigenvalue, and $n(\omega)$ is the spectral density
~\cite{SB,generalize}.  The eigenvalue $\omega_{min}$ is related to large
scale properties of the network, whereas $\omega_{max}$ is related to 
local, non-universal properties of the network ~\cite{SB}.  If we take the
network to be fractal, then $n(\omega) \sim \omega^{(d_s -2)/2}$, where 
$d_s$ represents the effective spectral dimension of the network ~\cite{AO}.
An unconstrained ideal polymer has $d_s =1$ and $d_s$ increases with the 
number of constraints.
For $d_s > 2$, $R_G^2$ depends on $\omega_{max}$, and therefore on the local
properties of the network.
For $d_s <2$, $\omega_{min} \sim N^{-2/d_s}$, 
so $R_G^2 \sim \omega_{min}^{(d_s - 2)/2} \sim N^{(2-d_s)/d_s}$ ~\cite{SB}.
The transition between these two regimes occurs when $d_s = 2$ and therefore
$R_G^2 \sim \log N$, which is precisely the dividing line obtained from 
using our localization criterion with $R_G^2 \propto N/M$.
Although this derivation approximates the constrained chain as a homogeneous
fractal, it definitely shows that, if the results of KK 
and of Solf and Vilgis ~\cite{SV} are correct for our model,
then the localization criterion is
related to the behavior of general Gaussian networks
in an intriguing and unexpected way.

We thank Jack Douglas, S. Havlin, and 
T. R. Kirkpatrick
for useful discussions.  
This research was supported 
by the National Institutes of Health through the Division of Computer Research
and Technology and the National Science Foundation through Grant
Number NSF CHE 93-0675.


\begin{references}
\bibitem{generalize}
In their paper, Sommer and Blumem only consider the case where all of the 
bonds have the same spring constant; however, the generalization to 
systems with different spring constants is trivial and does not affect 
the quoted results in any substantial way.
\end{references}
\end{document}